# EXACT RESULTS OF THE ONE–DIMENSIONAL TRANSVERSE ISING MODEL IN AN EXTERNAL LONGITUDINAL MAGNETIC FIELD

Jozef STREČKA, Hana ČENČARIKOVÁ, Michal JAŠČUR
Department of Theoretical Physics and Astrophysics, Institute of Physics, P. J. Šafárik University, Moyzesova 16, 041 54 Košice, Slovak Republic, tel. 055/6222121–23, E-mail: jozkos@pobox.sk, jascur@kosice.upjs.sk

**SUMMARY**

*Magnetic properties of the 1D mixed spin–1/2 and spin–S (S >1/2) transverse Ising model in the presence of an external longitudinal magnetic field are calculated exactly by the use of the generalised decoration–iteration mapping transformation. By assuming that only the spin–S atoms do interact with the transverse field the exact results for the Gibbs free energy, longitudinal magnetization, transverse magnetization, entropy and specific heat have been derived and discussed in detail. In addition to the standard temperature variations of the transverse magnetization, another two non–trivial thermal dependences have been found. The results support the concept that the spin reorientation from the longitudinal towards the transverse field direction takes place in the narrow temperature region.*

## 1. INDRODUCTION

Extensive theoretical as well as experimental work has been done over the years to explore the behaviour of the pure regular magnetic chains. In particular, an experimental discovery of a large variety of bimetallic chains has stimulated the renewal interest in studying these systems. As a matter of fact, the most of synthesized magnetic chains usually consist of two regularly alternating magnetic ions with unequal spin magnitudes that form two unequal sublattice of the magnetic chain. The strong interest in these systems arises as a consequence of the fact that the mixed–spin chains show many outstanding features, such as the magnetization plateau [1], the Kostrelitz–Thouless transition [2] or the double peak specific heat curve [3]. An important class of the mixed–spin chains constitute the spin–1/2 and spin–*S* (*S* >1/2) magnetic chains, like $A^{II}Cu^{II}(pbaOH)(H_2O)_3.2H_2O$ [4], where $Cu^{II}$ is a spin–1/2 magnetic ion and $A^{II}$ denotes one of the magnetic ions A = $Ni^{II}$ (*S* = 1), $Co^{II}$ (*S* = 3/2), $Fe^{II}$ (*S* = 2) or $Mn^{II}$ (*S* = 5/2) .

Recently, the considerable attention has been paid also to the investigation of the transverse Ising model. The main reason of such a great interest can be attributed to the fact that the transverse Ising model is very valuable model because of its different possible application. In fact, the transverse Ising model enables a simple explanation of the quantum properties of the hydrogen bonded ferroelectrics [5], cooperative Jahn–Teller systems [6], as well as strongly anisotropic magnetic materials in the transverse field [7]. More details about other possible application of the model can be found in an excellent reviews of Blinc and Zeks [8] and Stinchcombe [9]. Although, the transverse Ising model is the simplest quantum model, the complete exact solution have been obtained in the one–dimensional case only [10]. However, as far as we know, there has not been exactly examined the mutual effect of the longitudinal and transverse magnetic field on the magnetic properties of the mixed–spin chains, so far. Therefore, the main purpose of this work is to develop an accurate treatment of the transverse Ising model in the longitudinal magnetic field for the mixed spin–1/2 and spin–*S* (*S* >1/2) chains. Using the generalised mapping transformation technique we will treat the model exactly and within the framework of this method, the joint effect of the longitudinal and transverse magnetic field will be explored in detail.

The outline of the present paper is as follows. The fundamental framework of the transformation method is presented in Section 2. This is followed by a presentation of the most interesting numerical results in Section 3. Finally, some concluding remarks are given in Section 4.

## 2. FORMULATION

In this article we are concerned with the study of the mixed–spin chain composed of regularly alternating spins, the spin–1/2 (sublattice A) and spin–*S* (*S* >1/2, sublattice B) placed in the external longitudinal and transverse magnetic field. In general, the most difficult problem appearing in the quantum statistical models consists in the noncommutability of the relevant operators involved in the Hamiltonian. To overcome this difficulty, we consider only the special case of the system, when the transverse field acts just on one kind of atoms, namely, those with the spin variable *S* (sublattice B). Under this circumstance, the model is exactly solvable and the total Hamiltonian can be written as a sum of commuted bond Hamiltonians $H_k$

$$H = \sum_{k=1}^{N} H_k, \tag{1}$$

where $N$ is a total number of atoms of sublattice B and the bond Hamiltonian $H_k$ includes all the terms associated with $k$th spin variable $S_k$ of sublattice B

$$H_k = -S_k^z \left(E_k + H_B\right) - S_k^x \,\Omega - \frac{H_A}{2}\left(\mu_{k1} + \mu_{k2}\right), \qquad (2)$$
$$E_k = J\left(\mu_{k1} + \mu_{k2}\right).$$

In above, $\mu_{k\alpha}$ ($\alpha$=1,2) denotes the spins of sublattice A and $J$ represents the exchange constant between nearest–neighbouring atoms. The parameter $\Omega$ stands for the transverse field interaction of atoms of sublattice B and finally, the terms $H_A$ and $H_B$ describe the influence of an external longitudinal field on atoms of sublattice A and sublattice B, respectively.

To step forward with the calculation, one has to diagonalise the bond Hamiltonian $H_k$, what can be straightforwardly performed by making use of the well–known rotational transformation [11]. Subsequently, after an elementary algebra we can write following equation for the partition function $Z$ of the model under investigation

$$Z = \sum_{\{\mu\}} \prod_{k=1}^{N} \exp\left[\beta H_A \left(\mu_{k1} + \mu_{k2}\right)/2\right] \sum_{n=-S}^{+S} \cosh\left(\beta n \sqrt{\left(E_k + H_B\right)^2 + \Omega^2}\right). \qquad (3)$$

Here $\beta = 1/k_B T$, $k_B$ being Boltzmann constant, $T$ the absolute temperature and the symbol $\sum_{\{\mu\}}$ means a trace over a degree of freedom of sublattice $A$. The relevant expression for the partition function $Z$ of the mixed–spin chain indicates the possibility to utilize the decoration–iteration transformation, which has been originally introduced by Syozi [12] and later remarkably generalised by Fisher [13]. This transformation takes in our case the form

$$\exp\left[\beta H_A \left(\mu_{k1} + \mu_{k2}\right)/2\right] \sum_{n=-S}^{+S} \cosh\left(\beta n \sqrt{\left(E_k + H_B\right)^2 + \Omega^2}\right) = \qquad (4)$$
$$= A \exp\left[\beta R \mu_{k1} \mu_{k2} + \beta H_0 \left(\mu_{k1} + \mu_{k2}\right)/2\right].$$

Actually, this transformation provides an exact mapping relationship between the mixed–spin chain and its corresponding exactly solvable spin–1/2 Ising chain with an exchange integral $R$, in an external longitudinal field $H_0$. For the unknown parameters $A$, $R$ and $H_0$ that emerging in the decoration–iteration transformation one obtains after the standard procedure [12] following relations

$$A = \left(V_1 V_2 V_3^2\right)^{1/4},$$
$$\beta R = \ln\left(\frac{V_1 V_2}{V_3^2}\right),$$
$$\beta H_0 = \beta H_A + \ln\left(\frac{V_1}{V_2}\right). \qquad (5)$$

Here, we have defined the expressions $V_1$, $V_2$ and $V_3$ by this set of equations:

$$V_1 = \sum_{n=-S}^{+S} \cosh\left(\beta n \sqrt{\left(J + H_B\right)^2 + \Omega^2}\right),$$
$$V_2 = \sum_{n=-S}^{+S} \cosh\left(\beta n \sqrt{\left(J - H_B\right)^2 + \Omega^2}\right), \qquad (6)$$
$$V_3 = \sum_{n=-S}^{+S} \cosh\left(\beta n \sqrt{H_B^2 + \Omega^2}\right).$$

The substitution of the transformation (4) into the formula (3) for the partition function $Z$ leads immediately to the relationship between the partition function $Z$ of the mixed–spin chain and the partition function $Z_0$ of the standard spin–1/2 Ising chain

$$Z = A^N Z_0 \left(\beta R, H_0\right). \qquad (7)$$

This equality proves that the mapping relation (4) establishes an equivalence between the mixed–spin chain in the transverse field and the simple spin–1/2 Ising chain, since both partition functions differs from each other only by the multiplicative factor $A$ given by equation (5). Therefore, the equation (7) may be considered as a mathematical expression of the aforementioned exact mapping relationship between both models. However, the equality (7) is also very suitable for deriving other quantities, such as Gibbs free energy, internal energy, magnetization, entropy etc. Thus, for instance the Gibbs free energy $G$ of the mixed–spin chain is given with respect to the equality (7) by

$$G = -N R/4 - N \beta^{-1} \ln A - N \beta^{-1} \ln B, \qquad (8)$$
$$B = \cosh\left(\beta H_0/2\right) + \sqrt{\sinh^2\left(\beta H_0/2\right) + \exp\left(-\beta R\right)}.$$

At this stage, the magnetization can be easily calculated from the Gibbs free energy as a derivative with respect to the relevant magnetic fields. After straightforward, but a little bit tedious algebra, the longitudinal and transverse magnetization can be written in this compact form

$$m_A = \left\langle \mu_i^z \right\rangle = m_0,$$
$$m_B = \left\langle S_j^z \right\rangle = \left(K_1 - K_2\right)/2 + m_0 \left(K_1 + K_2\right) -$$
$$- \varepsilon_0 \left(K_1 - K_2 - 2K_3\right)/2,$$

$$m = (m_A + m_B)/2,$$
$$m^x = \langle S_j^x \rangle = (L_1 - L_2)/2 + m_0 (L_1 + L_2) - \quad (9)$$
$$- \varepsilon_0 (L_1 - L_2 - 2L_3)/2.$$

where $m^x$ denotes the transverse magnetization of the sublattice $B$ (the transverse magnetization of the sublattice $A$ is equal to zero), $m$ denotes the total longitudinal magnetization, $m_A$ and $m_B$ stand for the longitudinal magnetization of sublattice $A$ and sublattice $B$, respectively. The magnetization $m_0$ is the famous Ising result for the spin–1/2 linear chain

$$m_0 = \frac{1}{2} \frac{\sinh(\beta H_0 / 2)}{\sqrt{\sinh^2(\beta H_0 / 2) + \exp(-\beta R)}}, \quad (10)$$

and the term $\varepsilon_0$ is given by

$$\varepsilon_0 = \frac{4 m_0^2 \exp(-\beta R)}{\sinh^2(\beta H_0 / 2) + m_0 \sinh(\beta H_0)}. \quad (11)$$

Finally, the coefficients $K_1$, $K_2$, $K_3$ and $L_1$, $L_2$, $L_3$ appearing in the formulas (9) are given by this set of equations:

$$K_1 = F(J + H_B), \qquad L_1 = G(J + H_B),$$
$$K_2 = F(J - H_B), \qquad L_2 = G(J - H_B), \quad (12)$$
$$K_3 = F(H_B), \qquad L_3 = G(H_B),$$

whereby the function $F(x)$ and $G(y)$ are defined by

$$F(x) = \frac{x}{\sqrt{x^2 + \Omega^2}} \frac{\sum_{n=-S}^{+S} n \sinh\left(\beta n \sqrt{x^2 + \Omega^2}\right)}{\sum_{n=-S}^{+S} \cosh\left(\beta n \sqrt{x^2 + \Omega^2}\right)},$$
$$G(y) = \frac{\Omega}{\sqrt{y^2 + \Omega^2}} \frac{\sum_{n=-S}^{+S} n \sinh\left(\beta n \sqrt{y^2 + \Omega^2}\right)}{\sum_{n=-S}^{+S} \cosh\left(\beta n \sqrt{y^2 + \Omega^2}\right)}. \quad (13)$$

For completeness, we should mention that the entropy $S$ and specific heat $C$ of the studied system can be also very easily achieved from the Gibbs free energy using the basic thermodynamic formulas

$$S = -\left(\frac{\partial G}{\partial T}\right)_{H,\Omega} \quad \text{and} \quad C = -\left(\frac{\partial^2 G}{\partial T^2}\right)_{H,\Omega}. \quad (14)$$

## 3. NUMERICAL RESULTS AND DISCUSSION

Before discussing the most interesting numerical results it should be emphasized that the mixed–spin chains exhibit qualitatively completely identical behaviour independently of the spin value of sublattice B. Owing to this fact, we restrict our analysis only to two particular cases, namely, $S_B = 1$ and $S_B = 3/2$. Furthermore, we will consider for the sake of simplicity the same longitudinal field affecting both type of atoms, i. e. $H_A = H_B = H$.

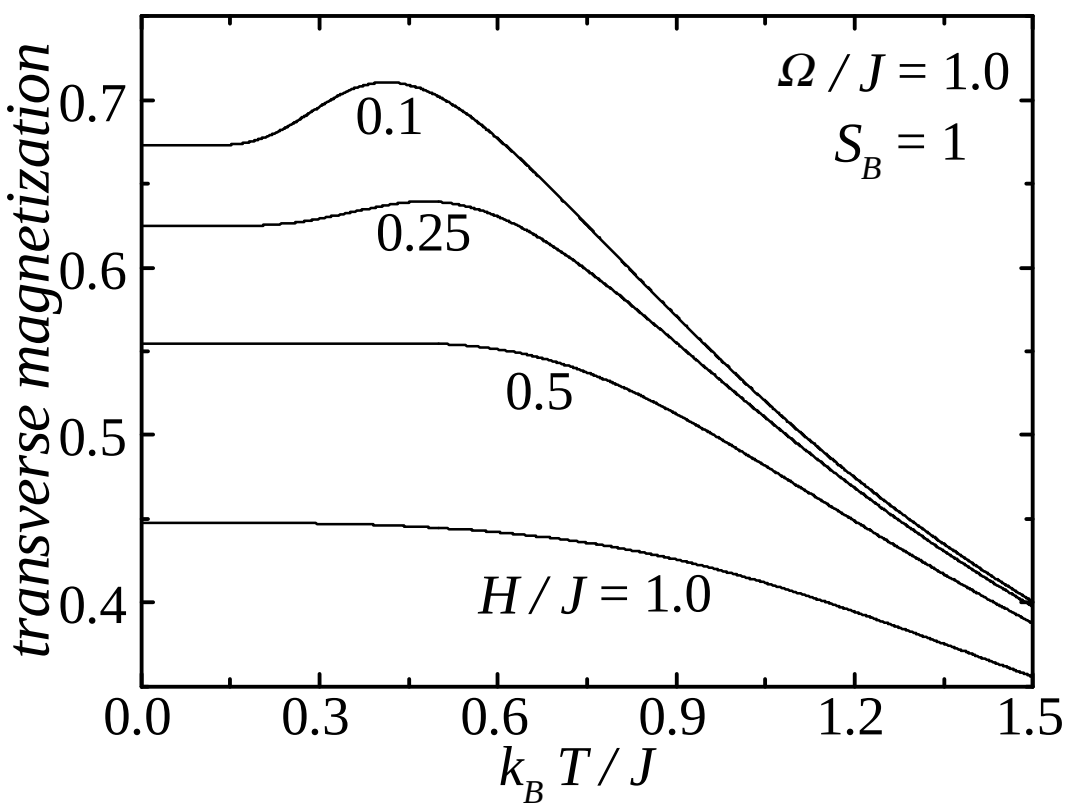

**Fig.1** The transverse magnetization against the temperature for the fixed transverse field and selected longitudinal fields when $S_B = 1$.

At first, the temperature variations of the transverse magnetization will be examined in detail. As one can observe from the Fig. 1 and 2, the transverse magnetization may not monotonically decrease with increasing temperature as could be expected. An interesting nontrivial dependence of the transverse magnetization, however, arises only for the sufficiently small longitudinal fields. Moreover, we should note that the character of the transverse magnetization curve depends above all on the transverse field strength. In fact, as shown in Fig. 1 the transverse magnetization by the strong transverse fields increases from its initial value, reaches a local maximum and then tends to zero with the increasing temperature. For better understanding of this behaviour, we have depicted in Fig. 3 both transverse as well as longitudinal magnetization thermal dependences. Interestingly, the thermal fluctuation may increase the transverse magnetization only in the small temperature region, where the longitudinal magnetization falls down very rapidly. These observation would suggest that the spins are in this small temperature region released from the longitudinal field direction and tending to align into the transverse field direction. Consequently, thermal induced increase of the transverse magnetization can be observed only in this small temperature region. On the other hand, as one can see from Fig. 2, the transverse magnetization for the smaller transverse field decreases from its initial value to a local minimum, then gradually increases in the narrow temperature region to reach a local maximum and finally sharply decreases as the temperature is raised. The other type of the magnetization curve is related to the fact that by the smaller transverse fields the appropriate fall in the longitudinal magnetization takes place at higher temperatures. Therefore, the transverse magnetization initially decreasing and only then increasing due to the thermal fluctuation. Obviously, the similar thermal induced increase in the transverse magnetization can not be found at higher longitudinal fields, since the corresponding fall in the longitudinal magnetization is smaller and is shifted towards the higher temperatures. As a consequence of that, thermally caused lowering of the transverse magnetization is in this temperature region greater than corresponding increase associated with the spin release from the longitudinal field direction.

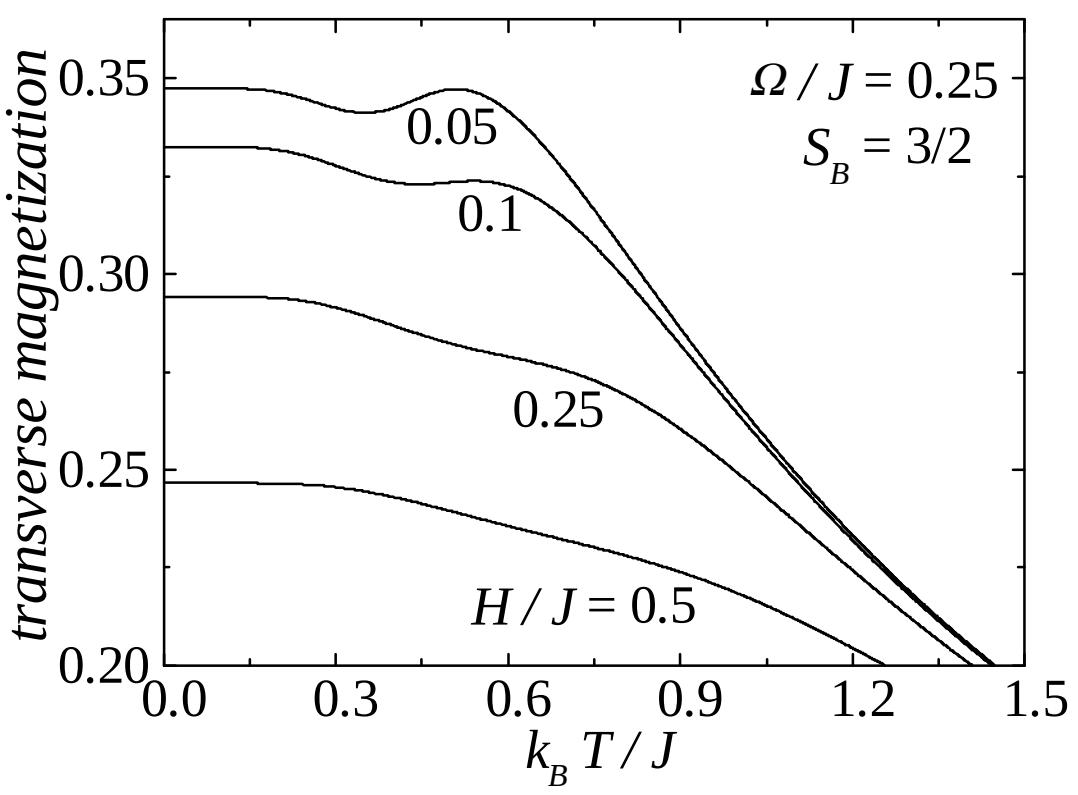

**Fig. 2** The thermal variations of the transverse magnetization for the fixed transverse field and selected longitudinal fields when $S_B = 3/2$.

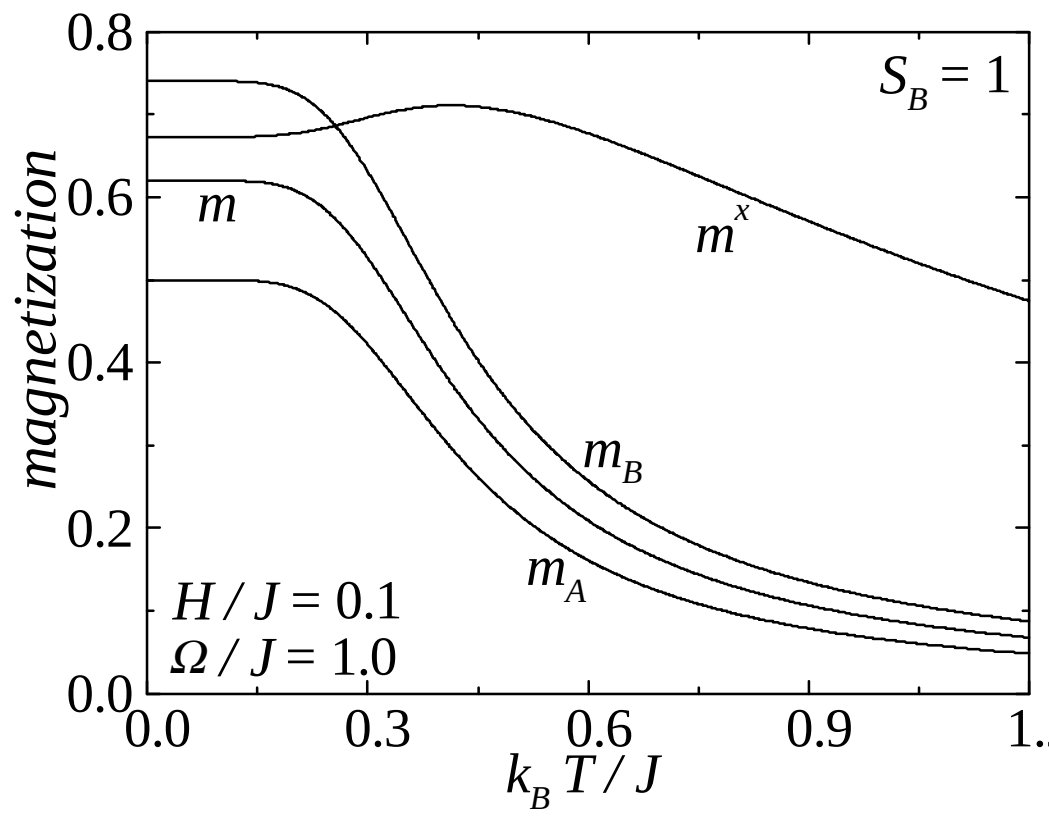

**Fig. 3** The total and sublattice longitudinal and transverse magnetization for fixed external fields as a function of temperature.

Now, let us turn back to the sublattice magnetization dependences on temperature, as shown in Fig. 3. Although, the longitudinal magnetization of both sublattices behave similarly, increasing of the transverse field leads to the disordering of the spins of sublattice B from the longitudinal field direction. On the other hand, the spins of sublattice A that are not directly affected by the transverse field, remain at the ground state perfectly ordered in the longitudinal field direction, regardless of the transverse field value. The previous results can be explained in the concept of Heisenberg uncertainty principle, namely, the longitudinal magnetization of the sublattice B is lowered by the transverse field on account of the raising transverse magnetization.

Finally, we will illustrate in Figs. 4 and 5 the thermal dependences of the entropy and specific heat. As one can see from the Fig. 4, the entropy exhibits the standard *S*–shaped curve independently of the transverse and longitudinal field strength. Nevertheless, one can ascertain that the curve becomes more abrupt and shifted towards the lower temperatures as the longitudinal field is lowered. Hence, the spin ordering (which occurs due to the effect of longitudinal field) clearly manifests itself in the broadening of the *S*–shaped curve. For better orientation, the corresponding temperature variations of the specific heat are plotted in Fig. 5. From the displayed dependences one can realize that the specific heat shows the familiar Schottky–type behaviour irrespective of the value of the external transverse and longitudinal field. It is noteworthy that the stronger the longitudinal field, the broader the Schottky–type anomaly occurring in the specific heat. Therefore, the observed behaviour implies that the longitudinal magnetic field is responsible for the strengthening of the short–range order. Indeed, the short–range order strength which is proportional to the maximum of the specific heat is shifted towards the higher temperatures with the longitudinal field increasing. Interestingly, the field induced double peak structure of the specific heat can not be observed for any combination of the transverse and longitudinal field. Thus, comparing our results with those [3] for the Heisenberg mixed–spin chains indicates that the doubly peak structure of the specific heat arises purely due to the quantum effects originating from the Heisenberg exchange interaction.

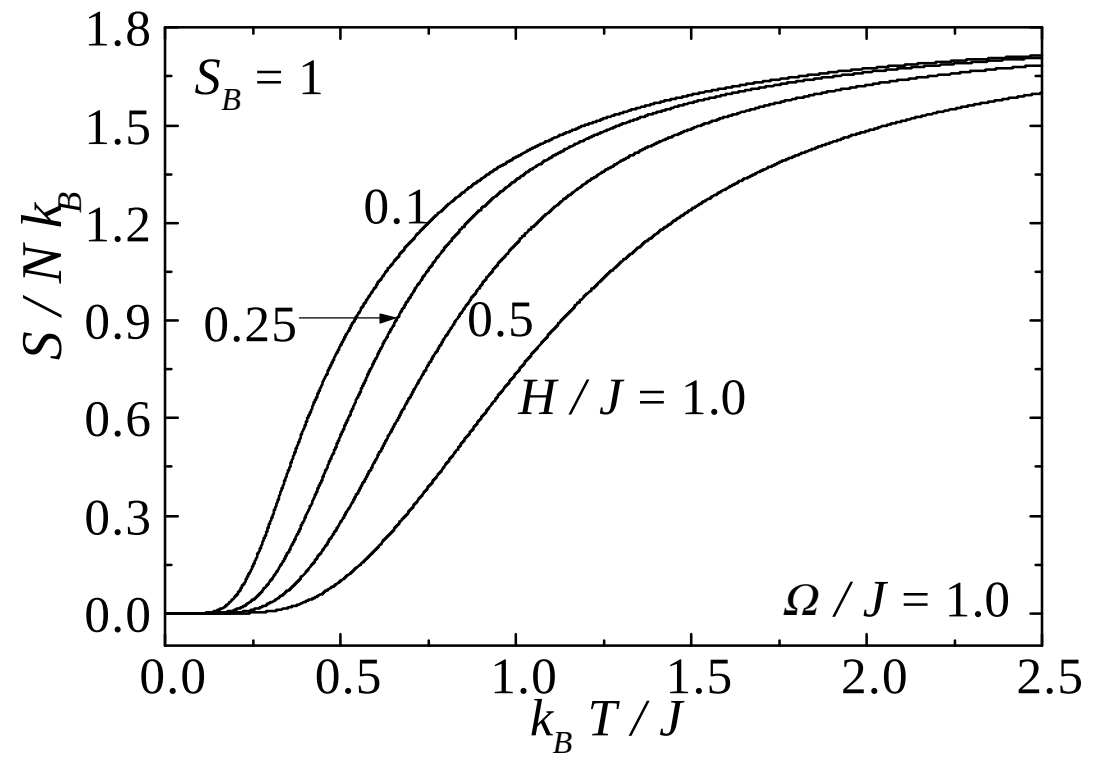

**Fig. 4** The entropy versus temperature for fixed transverse field and selected longitudinal fields.

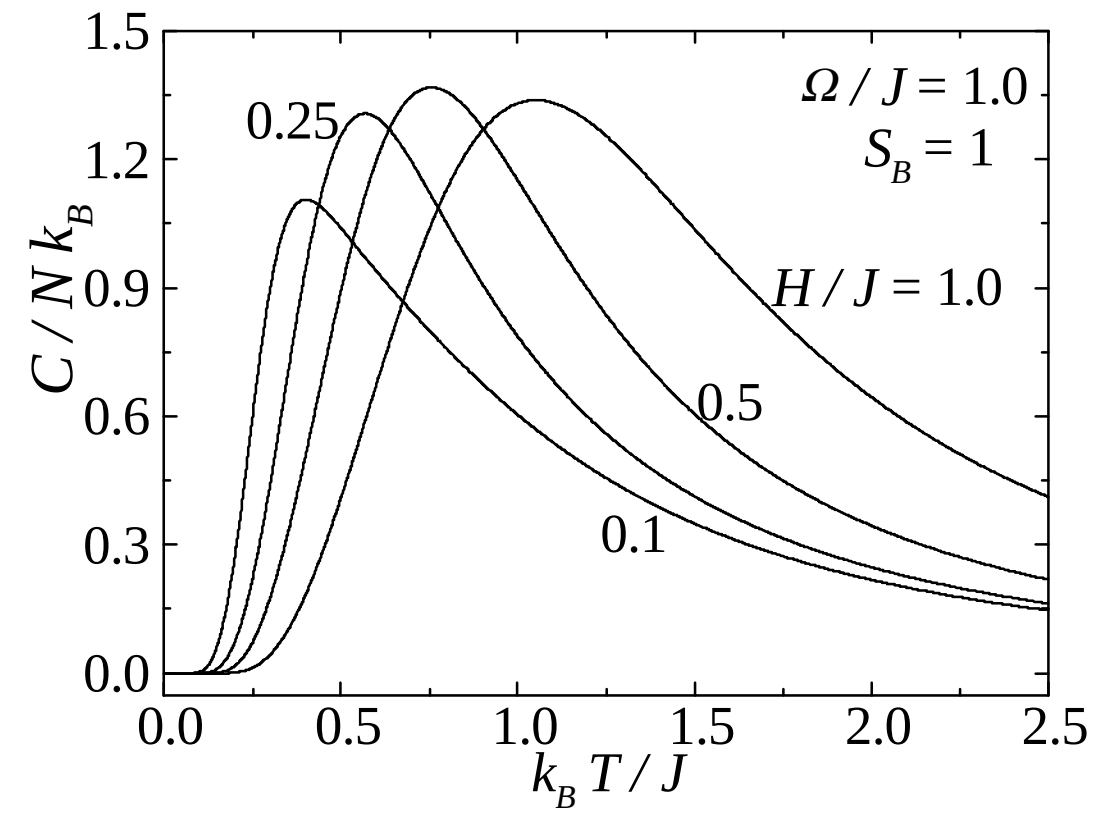

**Fig.5** The thermal dependence of the specific heat, when the longitudinal field is changed and transverse field is fixed.

## 4. CONCLUSION

In the present paper, by making use of an exact mapping relationship, the exact treatment for the mixed–spin chains in the presence of an external longitudinal and transverse field has been established. The presented results are interesting from the theoretical point of view because of their exactness, as well as from the experimental point of view in connection with many possible technological application of the one–dimensional magnetic materials, such as molecular based magnetic materials [14]. Perhaps, the most interesting result to emerge here is that there is a strong evidence for the spin release from the longitudinal towards the transverse field direction in the Ising–type anisotropic magnetic materials. We hope, that the theoretical prediction of the nontrivial thermal dependence of the transverse magnetization will inspire experimental physicists in order to confirm this phenomenon in the magnetism. Moreover, it should be emphasized that the applied transformation technique enables a simple calculation of the complete set of thermodynamic quantities. Altogether, the mapping technique provides a relative simple and simultaneously powerful tool for the investigation of the mixed quantum–classical systems. Actually, it turns out that it can be quite naturally extended to the more complicated quantum–classical system, such as the two–dimensional models in the transverse field [15] (of course, under the requirement of zero longitudinal field) or mixed Ising–Heisenberg systems [16]. In this way continues our next work.

*Acknowledgment*: This work has been supported by the Ministry of Education of Slovak Republic under VEGA grant No. 1/9034/02.

## BIOGRAPHY

Jozef Strečka was born on 23.4.1977. In 2000 he graduated with distinction at the Department of Theoretical Physics and Geophysics of the Faculty of Science at P. J. Šafárik University in Košice. At present, he is working as a PhD student at alma mater under the supervision of Associate Professor Michal Jaščur.
Hana Čenčariková was born on 17.12.1978. In 2002 she graduated with distinction at the Department of Theoretical Physics and Astrophysics of the Institute of Physical Science at P. J. Šafárik University in Košice.
Michal Jaščur was born on 16.10.1963. In 1987 he graduated (RNDr.) with distinction at the Department of Theoretical Physics and Geophysics of the Faculty of Science at P. J. Šafárik University in Košice. He defended his PhD at alma mater in 1995. From 2000 he holds the position of the Associate Professor.
The research interest of all the authors are related to the quantum theory of magnetism and phase transition phenomena.